# Radio frequency regenerative oscillations in monolithic high-$Q/V$ heterostructured photonic crystal cavities

Jinghui Yang[1*], Tingyi Gu[1*], Jiangjun Zheng[1], Mingbin Yu[2], Guo-Qiang Lo[2], Dim-Lee Kwong[2], Chee Wei Wong[1]

[1]*Optical Nanostructures Laboratory, Center for Integrated Science and Engineering, Solid-State Science and Engineering, and Mechanical Engineering, Columbia University, New York, NY 10027, USA*

[2]*The Institute of Microelectronics, 11 Science Park Road, Singapore Science Park II Singapore 117685, Singapore*

We report temporal and spectral domain observation of regenerative oscillation in monolithic silicon heterostructured photonic crystals cavities with high quality factor to mode volume ratios ($Q/V$). The results are interpreted by nonlinear coupled mode theory (CMT) tracking the dynamics of photon, free carrier population and temperature variations. We experimentally demonstrate effective tuning of the radio frequency (RF) tones by laser-cavity detuning and laser power levels, confirmed by the CMT simulations with sensitive input parameters.

* Electronic addresses: jy2484@columbia.edu and tg2342@columbia.edu

Regenerative oscillation, also termed self-induced oscillation or self-pulsation, is a cooperative temporal response between thermal and free carrier dispersions in an optical resonator, generating radio frequency (RF) tones with continuous wave (CW) input. Observations of regenerative oscillation in one dimensional (1D), 2D[1-8], 2.5D and 3D photonic resonators[9-14] are reported in various material structures. As silicon photonic crystals provides a complementary metal-oxide semiconductor (CMOS)-compatible platform, considerable studies on low energy switching and passive tuning are addressed in silicon-based devices[15-18], with potential in all-optical communication systems[19]. Monolithic photonic crystal cavities are characterized by their high quality factor to mode volume ratios ($Q/V$)[20-22], where optical thermal and free carrier bistability are demonstrated, enabled by strong light-matter interaction at sub-milliwatt power levels[15, 23] including pulsed carrier switching[24]. Here we report our observations of regenerative oscillation in a heterostructured cavity with quality factor ($Q$) ~ 500,000 and mode volume ($V$) ~ 0.11 $\mu m^3$ ($Q/V$ ratio of ~ $10^7$ $\mu m^{-3}$), with threshold of 79 $\mu W$ continuous wave input.

The double-/multi- heterostructured photonic crystal cavities are fabricated via 248 nm deep-UV photolithography and reactive ion etching on 250 nm-thickness silicon-on-insulator wafers. A scanning electron microscope (SEM) image in Fig.1a shows a typical double-heterostructure with a lattice constant ($a_1$) of 410 nm, air hole radii 0.276$a_1$, and increasing lattice constants of 415 nm and 420 nm ($a_2$) in the cavity region; the waveguides with length of 112$a_1$, width 1.0$a_1$ and 9 layers of air-holes separated from the cavity. 2 $\mu m$ oxide underneath the photonic crystal cavity is removed by buffered oxide wet etching. Inverse tapered couplers with an oxide over-cladding are integrated for optimized fiber-chip-fiber coupling. Inset of Fig.1a shows the 3D finite-difference time-domain (FDTD) simulations of the electric-field distribution, with the cavity modeled to have an intrinsic quality factor in excess of ~$10^6$ and modal volume $V_m$ of ~$1.2(\lambda_0/n)^3 = 0.11$ μm$^3$. To sustain oscillation in silicon photonic crystals, the threshold power of self-pulsation steadily increases with lower $Q/V$ ratio. For devices with $Q$ factor smaller than 100,000, regenerative oscillation is hardly observed even at 1 mW power levels.

Spectral transmission is used to characterize the high-$Q$ cavity, by sending the CW coherent transverse-electric light from a tunable laser onto chip through polarization controllers and piezoelectric-feedback-controlled lensed fibers. The average power level and temporal response of output light are monitored by power meter and fast photodetector (New Focus 1554B, 12 GHz bandwidth) respectively. The fast photodetector is connected to an electronic spectrum analyzer (Agilent 5052B) and a high-speed oscilloscope (Tektronix, 4 GHz bandwidth) to obtain the time-domain response of the high-$Q$ cavity. The environmental thermal fluctuation of the chip is minimized by a thermoelectric cooling module driven by a benchtop temperature controller (Thorlabs TED200C).

The measured optical transmission spectra with different input powers are shown in Fig.1b. Cavity resonance shifts from 1591.23 nm to 1591.39 nm as input power (all referred to the powers or estimated powers inside the cavity, same below) increases from 1 μW to 1 mW. By curve-fitting the transmission spectrum of cold cavity, we obtain the cavity linewidth of ~ 6 pm and 5 dB extinction ratio, corresponding to loaded and intrinsic $Q$ factors of 266,000 and 480,000 respectively. The $Q$ factors are slightly lowered by the imperfections in nanofabrication[25]. At higher input powers the cavity transmission becomes asymmetric due to thermal hysteresis effects, indicated by the sharp transition because of the bistable states[15, 26].

With the input power well beyond the threshold of bistability, temporal self-pulsation can be observed at a range of laser-cavity detunings. For an input CW laser power of 800μW, the output self-pulsation waveform is shown in Fig. 2a, at laser-cavity detunings of 120 pm, 190 pm and 267 pm respectively. The dependence of oscillation periods, dip widths (corresponding to the pulse width of cavity mode) and duty cycles (ratio between the dip width and the period) on laser-cavity detunings are systematically measured (Fig. 2b). The period of time domain pulsation is around tens of ns to 100 ns, with duty cycle varying from 20% to 60%. The upper and lower bounds of detuning in the oscillation region are illustrated in Fig 2c. Only within the bounded laser-cavity detuning ranges can the stable oscillation be observed, which displays decreased standard deviation of the time period of temporal pulses, along with much lower phase noise and amplitude noise compared to the oscillatory output in other ranges (e.g. chaotic states)[27]. Higher input power linearly increases the bound of detunings regions for the regenerative oscillations. For laser-cavity detunings smaller than 70 pm or larger than 220 pm, the oscillation is unstable and hard to measure.

Frequency domain response of the self-sustained regenerative oscillations is simultaneously monitored by the RF spectrum analyzer. The RF spectrum of the oscillation at the detuning of 160 pm is shown in Fig. 2d; the experimental relation between the fundamental RF tone and laser-cavity detuning is plotted in the inset of Fig. 2d, which are the Fourier transforms of the time-domain measurements (Fig. 2a). As detuning increases, initially chaotic excitation of the oscillation stabilizes, and more coherent oscillatory signals are generated, until the upper bound of the detuning. We also measured the single sideband phase noise and amplitude noise of the fundamental mode for regenerative oscillation. When stable temporal oscillation is generated, the amplitude noise decreases by ~ 40 dBc/Hz.

Fig. 3 shows a series of simulations of the temporal dynamics, phase diagram and spectrum under different input conditions. In the self-pulsation process, the dynamics of the cavity mode is modified by several nonlinear effects[4, 15, 28], i.e., thermo-optic effect, free-carriers dispersion and Kerr nonlinearity. The dispersion induced by Kerr effect is orders of magnitude weaker than the other two effects in silicon material devices. Competing resonance shifts therefore occurs – red shift caused by thermo-optic effect and blue shift by free-carriers dispersion, resulting in a modification of resonance wavelength temporally and periodically. We model the transmissions with time-domain nonlinear coupled mode theory[29, 30], where the dynamics of mode amplitude $a$ (square root of mode energy), the free carrier density $N$ and the cavity temperature shift $\Delta T$ are given by:

$$\frac{da}{dt} = (i(\omega_L - \omega_0 + \Delta\omega) - \frac{1}{2\tau_t})a + \kappa\sqrt{P_{in}} \tag{1}$$

$$\frac{dN}{dt} = \frac{1}{2\hbar\omega_0 \tau_{TPA}} \frac{V_{TPA}}{V_{FCA}^2} |a|^4 - \frac{N}{\tau_{fc}} \tag{2}$$

$$\frac{d\Delta T}{dt} = \frac{R_{th}}{\tau_{th}\tau_{FCA}} |a|^2 - \frac{\Delta T}{\tau_{th}} \tag{3}$$

where the total loss rate is $1/\tau_t = 1/\tau_{coup} + 1/\tau_{in} + 1/\tau_{TPA} + 1/\tau_{FCA}$. $1/\tau_{coup}$ and $1/\tau_{in}$ are the loss rates coupled into waveguide and into free-space respectively. The free carrier absorption rate $1/\tau_{FCA} = c\sigma N(t)/n$, and the two photon absorption rate is defined as $1/\tau_{TPA} = \beta_2 c^2/n^2/V_{TPA}|a|^2$, where $\beta_2$ is the effective two photon absorption coefficient of silicon. $V_{TPA}$ and $V_{FCA}$ are mode volumes of two photon absorption and free carrier absorption respectively. Based on the model, the cavity stored energy $U(fJ)$, free-carrier density $N(m^{-3})$ and temperature variation $\Delta T(K)$ are simulated on the left panels of Fig. 3; the phase diagrams of mode amplitude are shown on the middle panels; and the fast Fourier transform (FFT) spectra in the frequency domain on the right. The selected input conditions are: (a) input power 800 μW, detuning 160 pm; (b) input power 800 μW, detuning 30 pm; (c) input power 100 μW, detuning 30 pm, where (a) matches with our experimental measurements.

To match the experimental RF fundamental mode and parameter space of detuning (detailed in Fig. 4a and Fig. 4b), we adjust the initial conditions (I. C.) of internal cavity energy, free carrier density and temperature variation to be non-zero for larger detunings (e.g. larger than 50 pm at 800 μW input power). Mathematically this is analogous to the initial photons induced by thermo-optic effect before the pulsation is ignited. The non-zero I. C. in the model is derived iteratively in the range of temporal dynamics, e.g. in Fig. 3a, we set the initial condition of the mode energy to be 4 fJ, the free carrier density $N$ to be $4 \times 10^{21}$ m$^{-3}$ and the cavity temperature shift $\Delta T$ at 2.2 K; it matches well with the observed pulsation as shown in the RF spectrum in the inset of the right panel (the same as Fig. 2d). As a comparison, zero I. C. can sufficiently initiate the self-pulsation for detunings less than 50 pm from simulation; the relation between duty cycle and detuning matches experiment qualitatively. Fig. 3b shows the simulated self-pulsation at detuning of 30pm, with zero I. C., i.e. $U_0 = 1\times10^{-15}$ fJ, $N_0 = 1\times10^{19}$ m$^{-3}$(an estimated quiescent value for silicon), $\Delta T_0 = 0.1$ nK. Our experimental result, however, shows unstable or chaotic oscillation with detunings less than 60 pm as mentioned above, even though the self-pulsation in smaller or negative detunings are observed or predicted in other types of devices[7, 12, 31]. In frequency domain, condition (b) shows a higher harmonic level indicating a lower purity of signal, which suggests weaker device performance. The phase diagrams of (a) and (b) are similar, both with complex limit cycles. At a lower input power such as in the condition (c) (Fig. 3c), a purer signal is more likely to be excited, with only one limit cycle and lower harmonics; however, it is not observed experimentally possibly due to the low intensity or signal-to-noise ratios compared to condition (a).

We point out that, high frequency oscillations up to ~ 2 GHz can occur as shown in the simulation of the temporal dynamics and FFT spectrum of Fig. 3a, with decreased damping through improved thermal conductivity and cavity stored energy (as shown). Experimentally this is not observed due to either comparatively lower thermal conductivity in the device cavities or the limited data acquisition bandwidth. Similar phenomena are observed and explained in microdisk resonators[2], and silicon photonic crystal cavities[12] where material surface treatment is introduced to obtain non-damped and sinusoidal GHz oscillations.

Following the initial condition treatment as mentioned above, the parameter space (input power and detuning) and RF frequency of the fundamental mode are obtained through simulation by finely adjusting parameters in the model, e.g. free carrier lifetimes and mode volumes. The results are shown in Fig. 4a and Fig. 4b with experimental data as comparisons respectively. Fig. 4c depicts the simulated threshold power and cavity stored energy with different $Q$ factors, which illustrates that regenerative oscillation will only occur in high-$Q$ devices, with the threshold enhanced by the wavelength-scale photonic crystal modal volumes. The threshold $Q$ factor for observing the oscillation in silicon based photonic crystals is about 110,000, at 400 μW power level. Experimental data is highlighted, where the internally stored cavity energy is estimated to be 5 fJ [32].

In order to accurately estimate the free carrier lifetime, which is a function of free carrier density [2, 4], we plot the dependence of the RF fundamental mode versus carrier lifetime, and map the experimental measurement in Fig. 4d, where we find 0.5 ns as the best fit for experiments. The free carrier lifetime can be electro-optically controlled by placing lateral bias on the high $Q$ cavity, for a potential voltage controlled local oscillator towards on-chip signal processing.

In this work we report our observations of self-sustained regenerative pulsation in photonic crystal heterostructured cavities with $Q/V$ ratio of $10^7$ μm$^{-3}$, with detailed measurements and numerical interpretations. As the input power into the cavity increases up to ~ 80 μW (internal stored energy up to ~ 5 fJ), the transmission intensity self-excites into an oscillatory behavior in the time-domain. The period of time domain pulsation is around 100 ns, with duty cycle varying from 20% to 60%, controllable by the input powers and detunings. The oscillation region is mapped onto various laser-cavity detunings versus coupled powers. The complete nonlinear coupled mode theory simulation illustrates the parameter space, and specifies the effective free carrier lifetime in the cavity. The comparison between experiments and simulations under different input conditions is described and discussed. The phenomena and studies allow lower power operation for all-optical signal processing and yield an alternative approach towards tunable frequency oscillators by controlling the incident drive signal.

**Acknowledgement**

The authors acknowledge helpful discussions with James F. McMillan, and support from the ONR program on Nonlinear Dynamics, and the National Science Foundation IGERT (DGE-1069240).

**Figure and captions**

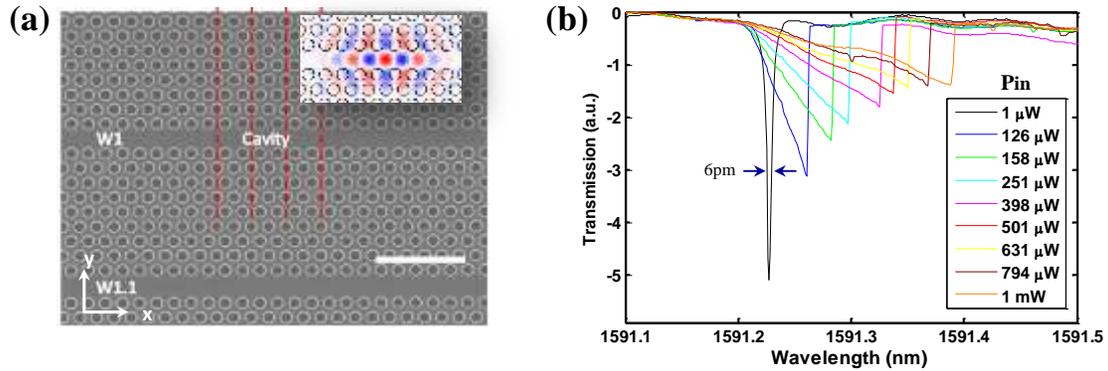

**Figure 1. Ultrahigh-*Q/V* platform for self-induced regenerative oscillations.** (a) Scanning electron micrograph of a double-heterostructured high-*Q* photonic crystal cavity (scale bar: 2 μm); (inset) 3D finite-difference time-domain calculated electric field profile ($E_y$) of the high-*Q* mode supported by a double-heterostructured cavity ($Q_{in}$ = 480,000, mode volume = 0.11 μm$^3$). (b) Measured transmission spectra (wavelength

scanned from low to high) with normalized input power from 1 µW to 1 mW. The linear transmission is measured at 1 µW (black symmetric lineshape).

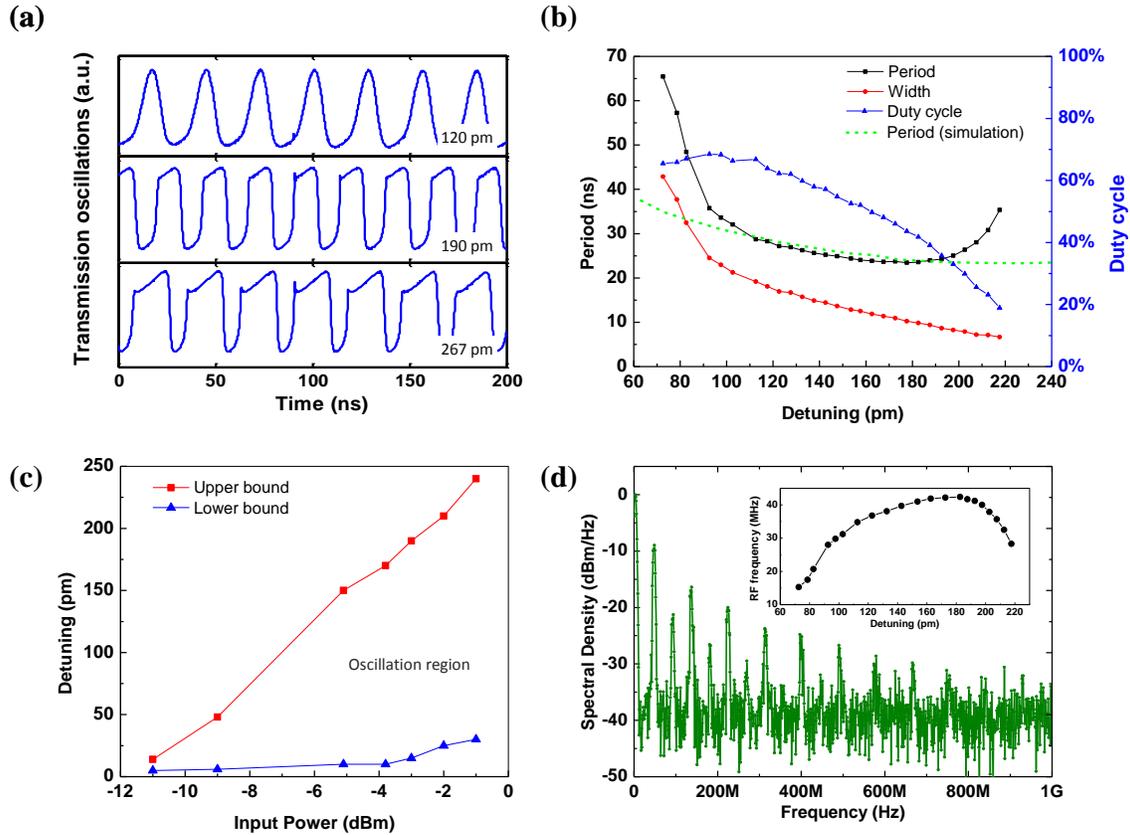

**Figure 2. Experimental observation of regenerative oscillations in a high-$Q$ double-heterostructured photonic crystal cavity, controlled by input laser power and laser-cavity detuning.** (a) Observed time domain periodic pulsation with laser-cavity detunings of 120 pm, 190 pm and 267 pm at 800 µW input power. (b) Period, dip width, and duty cycle versus detuning, summarized from a set of temporal observations as in panel (a). (c) Measured upper and lower bound of detuning range for oscillations with $Q$ = 480,000. (d) RF spectrum at the wavelength detuning of 163 pm; (inset) measured fundamental RF frequency versus detuning at 800 µW input power.

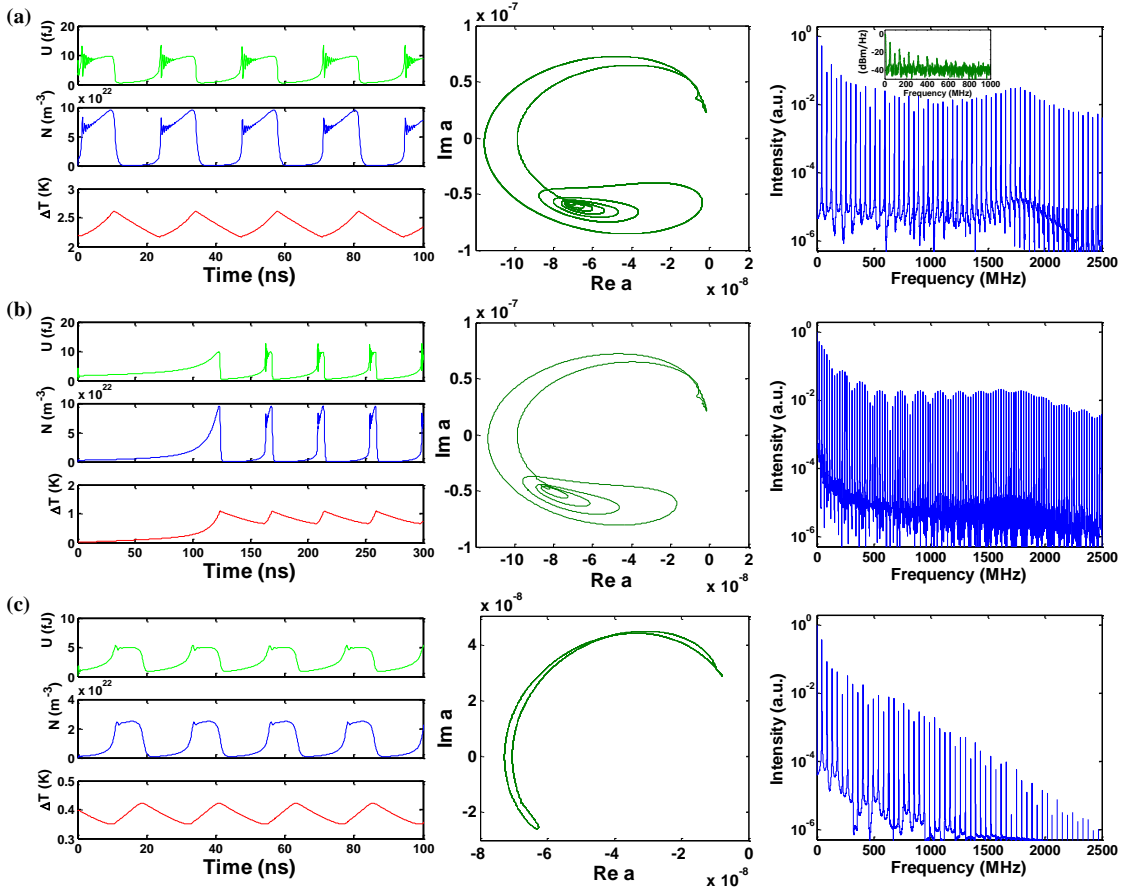

**Figure 3. Coupled mode theory simulations of the temporal dynamics (cavity stored energy *U(fJ)*, free-carrier density *N(m$^{-3}$)*, and temperature variation *ΔT(K)*), phase-space reconstruction of normalized mode amplitude, and frequency domain spectrum respectively under selected input conditions:** (a) input power 800 μW, detuning 160 pm, non-zero initial conditions (I. C.) ($U_0$ = 4 fJ, $N_0$ = 4×10$^{21}$ m$^{-3}$, $\Delta T_0$ = 2.2 K); (b) input power 800 μW, detuning 30 pm, zero I. C. ($U_0$ = 1×10$^{-15}$ fJ, $N_0$ = 1×10$^{19}$ m$^{-3}$, $\Delta T_0$ = 0.1 nK); (c) input power 100 μW, detuning 30 pm, non-zero I. C. ($U_0$ = 1 fJ, $N_0$ = 5×10$^{20}$ m$^{-3}$, $\Delta T_0$ = 0.4 K). (a) corresponds to our experimental conditions; (inset of right panel of (a)) the measured RF spectrum, which matches well with the simulation.

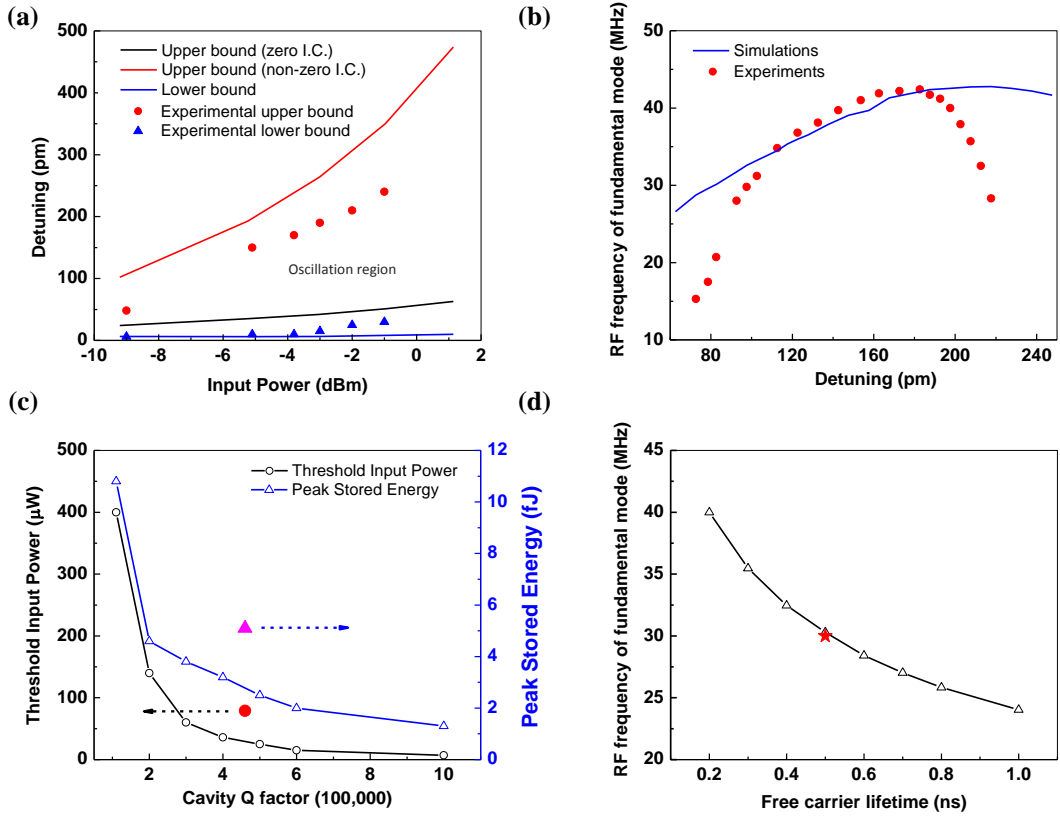

**Figure 4. Comparison of simulation and experiment.** (a) Simulated up and low bound of detuning range for oscillations for $Q = 480,000$. Red solid and dashed lines show the upper bound with non-zero ($U = 4$ fJ, $N = 4\times10^{21}$ m$^{-3}$, $\Delta T = 2.2$ K) and zero ($U = 1\times10^{-22}$ fJ, $N = 1\times10^{-10}$ m$^{-3}$, $\Delta T = 0.1$ nK) initial condition (I.C.) respectively, corresponding to the cases for the cavity with or without initial photons before the nonlinear behavior occurs. Experimental results (red and blue dots) are shown as a comparison with simulation. (b) Simulated (line) and measured (dots) RF frequency of fundamental mode of self-pulsation versus detuning. (c) Threshold input power (black) and peak intracavity energy (blue) as functions of cavity $Q$ factor. Experimental results marked as a solid dot/triangle. The threshold power for oscillation is beyond 1 mW in the cavity with $Q < 100,000$. (d) Free carrier lifetime dependent oscillation features characterized by the generated

fundamental RF frequencies with laser-cavity detuning at 10 pm. At given conditions, the measured RF frequency (red star) indicates the free carrier lifetime in the high $Q$ cavity to be 500 ps.